\documentstyle[12pt]{article}
\title{Static elastic deformations in general relativity}

\author{Llu\'{\i}s Bel\\ 
\it Laboratoire de Gravitation et Cosmologie Relativistes\\
\it CNRS/URA 769, Universit\'e Pierre et Marie Curie\\
\it 4, place Jussieu. Tour 22-12. Bo\^{\i}te courrier 142\\
\it 75252 PARIS Cedex 05, France}
\date{}
\begin{document}

\maketitle

\begin{abstract} 

We present a new approach to the theory of static deformations of elastic
test bodies in general relativity based on a generalization of the concept of
frame of reference which we identify with the concept of quo-harmonic
congruence. We argue on the basis of this new approach that weak
gravitational plane waves do not couple to elastic bodies and therefore the
latter, whatever their shape, are not suitable antennas to detect them.

\end{abstract}

\section {Introduction}

The theory of the behavior of resonant elastic test bodies in the field of
weak gravitational waves is based usually on the following approach: first
of all one uses a Fermi frame of reference with origin the center of mass of
the bar to derive a field of forces acting on the bar, or equivalently one
invokes the equation of geodesic deviation between the geodesic of the
center of mass and that of any other element of the bar; secondly one claims
that since the force fields derived using either one of the preceding
methods is very small one can use classical elasticity theory to describe
the coupling between the wave and the bar. We claim that the first idea on
which this approach is based is wrong, and the second needs to be understood
in a more general context. Our opinion is based on the following points: 

i) The identification of a Fermi congruence with a frame of reference is
unacceptable except in the case where the Fermi congruence is homogeneous,
i.e. it is such that any of the world-lines of the congruence can be
considered to be the origin of the congruence, in which case the Fermi
congruence is a Born rigid congruence.  This is not the case with the
congruence that is used to derive the field of forces in a resonant bar.
Moreover the justification of both, the Fermi congruences and the equation
of geodesic deviation, depends crucially on the implicit postulate according
to which whatever the circumstances being considered the quantity:

\begin{equation} \int{\sqrt{g_{\alpha\beta}dx^\alpha
\label {1.1} 
dx^\beta}} 
\end{equation} 
calculated along any space-like geodesic, between two events, one of them
lying on a world-line to this geodesic, can be identified to a physical
length.	This length-postulate, say, seems justified as being perfectly
symmetric to the time-postulate according to which the integral (\ref{1.1})
calculated between two events on any  time-like world-line can be identified
with a time interval measured with an appropriate clock. Besides their {\it
a priori} esthetic value both the time and length postulates have been
verified experimentally to some accuracy.

The time-postulate has been satisfactorily tested directly, it gives an
explanation of the dependence of the mean-life of elementary particles on
its velocity with respect a galilean frame of reference and it is daily at
work, so to speak, in navigation systems. Since the length-postulate is
equivalent to assuming the isotropy of the speed of light and its
independence of the event where it is measured, every experiment of the
Michelson-Morley type tests  it to some accuracy for small values of the
integral (\ref{1.1}). The best accuracy claimed up to date corresponds to an
anisotropy of less or of the order of $10^{-13}$ on the surface of the earth
at the location where the latest experiment was performed. \,\footnote{See
\cite{BH}. The value of $10^{-15}$ mentioned in this paper is the result of
some corrections that should not be made if one accepts that the observed
effect could have a local origin due, say, to the earth rotation, instead of
deciding that no effect could be real unless it is of cosmic origin}

So, both postulates are formally symmetric and both can be checked
experimentally to a high degree of accuracy. But there is an important
difference. The time postulate does not bring about any theoretical
difficulties and instead to accept the length-postulate as a law satisfied
with infinite precision leads to the conclusion that the only frames of
reference in any given space-time are the Born congruences. Since the latter
are rather scarce, even in special relativity, this point of view leaves
special and general relativity without a satisfactory theory of frames of
reference. We prefer to accept the necessity of having such a theory and to
examine the implications of this other point of view.\,\footnote{In
\cite{BMM} we discussed the implications concerning the speed of light using
a theory of frames of reference which is an ancestor of that summarized in
section 2}

ii) On the other hand the classical theory of elasticity is based on some
specific laws, like for instance Hooke's law, but above that it is based on
some general concepts common to every branch of classical mechanics like for
instance the concept of frames of reference which are the motions of
idealized objects called rigid bodies. Let us just remind for example that
the concept of deformation does not make sense until one has defined an
idealized standard, i.e. a frame of reference which by assumption, to a
desired approximation, has not been deformed. Therefore to invoke the
classical theory of elasticity in a problem of general relativity, whatever
small is the gravitational field which is being considered, should require
at least that we know what are the congruences which at the weak field limit
coincide with the rigid motions of classical mechanics, and this supposes of
course that a theory of the frames of reference has been developed in
general relativity. We must realize also that a gravitational wave, whatever
its field strength, does not have a limiting Newtonian equivalent and
therefore to deal with gravitational waves one has to be consistent also
with those other branches of physics that one uses to describe them.

Section 2 is a short review of the definition of quo-harmonic congruences
that we also call meta-rigid motions to emphasize our choice of interpreting
them as frames of reference. Quo-harmonic congruences are a generalization
of Born congruences and have already been discussed in two preceding papers
\cite{Salas}, \cite{Meta}.  Section 3 contains the presentation of our point
of view on the basic principles of the relativistic theory of elasticity.
The ingredients in our theory, some common to other approaches and some new,
are the following:

\begin{itemize}
\item A definition of the generating vector field of a deformation.
\item The strain tensor field defining a
deformation understood as relative to a given frame of reference.  
\item The energy-momentum tensor
\item The Hooke's law, and
\item The Beltrami-Michell integrability conditions. 
\end{itemize}

We shall consider only static deformations. These are defined as those
deformations for which the motion of the deformed body can be described by
the same congruence as that of the body before deformation.  In section 4 we
work out explicitly, as an example which illustrates the role of each of the
ingredients above, the special relativistic generalization of a bar being
accelerated with constant acceleration.  Section 5 contains an analysis of
the problem of the detection of weak plane gravitational waves. Is this
analysis which has led us to claim that such waves do not couple to
elastic bodies as we stated in the abstract.  

\section{Frames of reference}

Frames of reference must be a class of congruences with two main
properties, besides some more specific ones: 

i) its local definition must be intrinsic and independent of any particular
space-time. On the contrary global conditions compatible with the local
definition may be dependent on the space-time being considered. 

ii) any 3-dimensional sub-bundle of world-lines of a frame of reference must
caracterize the whole congruence, and satisfy the same local intrinsic
condition. 

Among acceptable congruences to be used in General relativity as frames of
reference are Born motions, \cite{Born}, when they exist but their number is
notoriously small, whatever the space-time being considered, including
Minkowski space-time, \cite{Herglotz}, \cite{Noether}.

Fermi congruences, either in their original form, \cite{Fermi},
\cite{Manasse}, or including rotation of the axis on the world-line which
generates the congruence, \cite{Mast}, \cite{NiZim}, are often used as a
substitute to Born congruences. This is in our opinion a serious mistake.
Fermi congruences obviously satisfy condition i) above, but they only
satisfy condition ii), and are therefore acceptable as frames of reference,
when they are homogeneous, i.e., when any of their world-lines could be
considered as a generator of the same congruence. This implies (See for
instance \cite{Salas}) that a Fermi congruence is in fact a Born congruence,
and we are again led to the difficulties stated in the preceding paragraph. 

The problem of generalizing the class of Born congruences has already been
considered in many papers. Born himself, \cite{Born2}, proposed a second
definition and since then many other generalizations have been proposed (See
for example more references in \cite{Bel0}). The definition of a frame of
reference to be used below has been derived from tentatives proposed in
several preceding papers \cite{Bel0}, \cite{ABMM}, \cite{BMM}, \cite{Salas},
\cite{Meta}.

Let us consider a time-like congruence $\cal R$ and let $u^\alpha$ be its
tangent unit vector field. We shall use the following definitions and
notations:

\begin{equation}
\label {2.16}
\hat g_{\alpha\beta} \equiv g_{\alpha\beta}+u_\alpha u_\beta
\end{equation}
which is the projector associated to $u^\alpha$,

\begin{equation}
\label {2.9}
\Lambda^\alpha \equiv - u^\rho \nabla_\rho u^\alpha
\end{equation}
which is, up to the sign, the intrinsic acceleration field, and	we shall
call occasionally the Newtonian Field. And

\begin{equation}
\label {2.18}
\Sigma_{\alpha\beta} \equiv c(\hat\nabla_\alpha u_\beta
+\hat\nabla_\beta u_\alpha)
\end{equation} 
where, $T_{\beta_1 \beta_2 \cdots \beta_s}$ being any tensor,

\begin{equation}
\label {2.23}
\hat\nabla_\alpha T_{\beta_1 \beta_2 \cdots \beta_s} \equiv
\hat g^\lambda_\alpha 
\hat g^{\mu_1}_{\beta_1}\hat g^{\mu_2}_{\beta_2}
\cdots \hat g^{\mu_s}_{\beta_s}\nabla_\lambda
T_{\mu_1 \mu_2 \cdots \mu_s}, 
\end{equation}
which is called the rate of `deformation' field\,\footnote{We write the word
deformation in quotation marks here to avoid a confusion with the concept of
deformation to be defined in the next section}.

Using in particular a system of coordinates adapted to $\cal R$, i.e., such
that $u^i=0$ the metric of the space-time can be conveniently described in
the following form:

\begin{equation}
\label {2.1}
ds^2 = -(\vartheta^0)^2 + d\hat s^2
\end{equation} 	 
with

\begin{equation}
\label {2.2}
\vartheta^0 = \xi(dx^0 - \varphi_i dx^i), \quad 
d\hat s^2 = \hat g_{ij}(t,x^k)dx^i dx^j, \quad  x^0=ct
\end{equation}
where the following notations have been used

\begin{equation}
\label {2.3}
\xi = \sqrt{-g_{00}} \quad , \quad \varphi_i = \xi^{-2}g_{0i} ,
\end{equation}
and

\begin{equation}
\label {2.15}
\hat g_{ij}=g_{ij}+\xi^2 \varphi_i\varphi_j
\end{equation}
which we shall call the Fermat quo-tensor\,\footnote{The prefix quo-
indicates that it is a tensor orthogonal to the vector $u^\alpha$ written in
a system of coordinates adapted to this vector} and it is the object whose
components are the space components of the projector (\ref{2.16}). The
quo-tensors whose components are the space components of $\Lambda^\alpha$ 
and $\Sigma_{\alpha\beta}$ are

\begin{equation}
\label {2.19}
\Lambda_i = -c^2(\hat\partial_i ln\xi +\partial_0 \varphi_i)
\end{equation} 
where

\begin{equation}
\label {2.5}
\hat\partial_i\cdot \equiv
\partial_i\cdot + \varphi_i\partial_0\cdot
\end{equation}
and

\begin{equation}
\label {2.20}
\Sigma_{ij}= c\hat\partial_0 \hat g_{ij}
\end{equation}
where

\begin{equation}
\label {2.21}
\hat\partial_0 \equiv \xi^{-1}\partial_0
\end{equation}

We shall also need to consider the following symbols, first used in
\cite{Zel'manov} and in \cite{Cattaneo}:

\begin{equation}
\label {2.4}
\hat\Gamma^i_{jk} = \frac12 \hat g^{is}(\tilde\partial_j\hat g_{ks}
+ \tilde\partial_k\hat g_{js} -\tilde\partial_s\hat g_{jk}) ,
\end{equation}

We shall say that a function $f(x^\alpha)$ is a quo-harmonic function in the
quotient space ${\cal V}_3={\cal V}_4/R$, where $R$ here is the equivalence
relation  defined by the congruence $\cal R$, if it is a solution of the
following equation

\begin{equation}
\label {2.6}
\hat\triangle f-\hat\partial_0^2 f=0
\end{equation}
where

\begin{equation}
\label {2.7}
\hat\triangle \equiv \hat g^{ij}(\hat\partial_i\hat\partial_j 
- \hat\Gamma^k_{ij}\hat\partial_k) 
\end{equation}

This quo-harmonicity concept is an intrinsic concept associated to a
congruence. In fact, using a general system of coordinates, eq. (\ref{2.6})
can be written in a manifestly covariant way as follows

\begin{equation}
\label {2.8}
\Box f + (\Lambda^\alpha + \Sigma u^\alpha)\partial_\alpha f=0
\end{equation} 
where:

\begin{equation}
\label {2.10}
\Box \equiv g^{\alpha\beta}(\partial_{\alpha\beta} -
\Gamma^\rho_{\alpha\beta}\partial_\rho)
\end{equation}
is the intrinsic d'Alembertian of the space-time and

\begin{equation}
\label {2.22}
\Sigma=\hat g^{\alpha\beta}\Sigma_{\alpha\beta}
\end{equation}.

We shall say that a congruence $\cal R$ is quo-harmonic if there exist three
independent functions of the space coordinates only, $f^a(x^i)$, which are
quo-harmonic, i.e. such that:

\begin{equation}
\label {2.11}
\hat\triangle f^a =0	
\end{equation}
Born congruences are quo-harmonic. This follows from the fact that for Born
congruences the Fermat tensor $\hat g_{ij}(x^k)$ does not depend on time and
therefore the operator $\hat\triangle$ above becomes the usual 3-dimensional
Laplacian in a Riemannian manifold. Eqs. (\ref{2.11}) admit then in the
neighborhood of any point a system of three independent solutions (See for
example \cite{Bers}) and therefore quo-harmonic congruences are an
acceptable generalization of Born congruences.

Congruences in general, and quo-harmonic congruences in particular, can be
classified according to their order of genericity, a concept that was
defined in \cite{Meta} and indicates how general is the tensor
$\Sigma_{ij}$. Let us remind here only that the quo-harmonic congruences
with smallest genericity are the Born congruences, which have a genericity
2, that the congruences with genericity 3 are the shear-free ones, i.e.
those satisfying: 

\begin{equation}
\label {2.12}
\hat\partial_0 \hat g^{ij} + \frac13\Sigma\hat g^{ij} = 0
\end{equation}
and that those with genericity 6 are those for which there exist a function
$c_1$ such that:

\begin{equation}
\label {2.13}
\hat\partial^2_0 \hat g^{ij} + c_1\partial_0\hat g^{ij} = 0
\end{equation}

The current orthodoxy claims that general covariance allows to say that any
congruence can be considered to be an acceptable frame of reference, a point
of view that leaves this concept without any meaning. That this is not the
answer to the problem of defining the concept of frame of reference can be
argued as follows: In 2-dimensional space-times the number of degrees of
freedom of Born congruences is 1 like for rigid motions in classical 
mechanics with the same dimensions. Therefore in two dimensions one can have
both general covariance and a correct theory of frames of reference. In
3-dimensional space-times it can be conjectured on the basis of preliminary
results, \cite{Llosa}, that the set of congruences, call it ${\cal B}_2$,
that are Born or shear-free congruences have 3 degrees of freedom, again as
much as rigid motions in classical mechanics with the same dimensions. The
simple fact that this conjecture might be true is again an indication that
there is no conceptual contradiction between general covariance and the
possibility of having a more restrictive theory of frames of reference.  Our
point of view is that that this problem will be solved also for
4-dimensional space-times in one way or another. Quo-harmonic congruences,
which generalize Born congruences in 4 dimensions, and include all 
shear-free congruences in 3 dimensions, \cite{Llosa}, are for the time being
our best educated guess. 

\section{Basic concepts of relativistic elasticity theory}

Many different approaches to a theory of elasticity in general relativity
have been proposed in the literature (See for instance \cite{Rayner},
\cite{Papapetrou}, \cite{Hernandez}, \cite{CarterQuintana}, and  
\cite{Lamoureux} which contains a review of many proposals). Our approach
here differs from all other we know of by insisting that to make sense of
concepts such as the concept of deformation of a body one needs to have a
well defined concept of reference frame allowing to compare the initial
state of the body with the result of acting on it, either with universal
forces like the gravitational ones or either with more specific ones. A
definition of this concept has been proposed in the preceding section and it
is used in this one in an essential way in conjunction with common
ingredients used to describe the behavior of macroscopic bodies like the
conservation of the energy-momentum tensor or a relativistic generalization
of Hooke's law.

Let us consider a system of coordinates adapted to a frame of reference
$\cal R$ and let us consider a second congruence $\cal B$ whose parametric
equations, $x^0$ being the parameter, in this system of coordinates are

\begin{equation}
\label {3.22}
y^\alpha = x^\alpha + \zeta^\alpha(x^\rho)
\end{equation}
If $\zeta^\alpha$ is a small vector so that the products of components can
be neglected we shall say that the congruence $\cal B$ is a deformation of
the frame of reference $\cal R$ and that $\zeta^\alpha$ is the generator of
this deformation. A short and explicit calculation proves that if we note
$w^\alpha$ the unit tangent vector to the congruence $\cal B$ and use the
following notation

\begin{equation}
\label {3.23}
w^\alpha = u^\alpha + \delta u^\alpha
\end{equation}
one has in this system of adapted coordinates:

\begin{equation}
\label {3.24}
\delta u^0 = -\xi^{-1} \varphi_i \partial_0 \zeta^i, \quad
\delta u^i = \xi^{-1}\partial_0 \zeta^i,
\end{equation}
a result which can be written in the manifestly covariant form:

\begin{equation}
\label {3.25}
\delta u^\alpha = \hat g^\alpha_\lambda {\cal L}(\zeta)u^\lambda  
\end{equation}

where $\cal L()$ is the Lie derivative operator. It follows from this
result that the generator of a deformation is defined up to the
transformation

\begin{equation}
\label {3.27}
\zeta^\alpha \rightarrow \zeta^\alpha + k u^\alpha
\end{equation}
where $k$ is any function. This property can thus be used to assume, without
any loss of generality, that $u^\alpha$ and $\zeta^\alpha$ are orthogonal:

\begin{equation}
\label {3.28}
u_\alpha \zeta^\alpha = 0
\end{equation} 

By definition the strain tensor of the congruence $\cal B$ with respect
to the frame of reference $\cal R$ is the tensor:

\begin{equation}
\label {3.26}
\epsilon_{\alpha\beta}=
\hat g^\lambda_\alpha\hat g^\mu_\beta{\cal L}(\zeta)\hat g_{\lambda\mu}
\hbox{    or   }
\epsilon_{\alpha\beta}= \hat\nabla_\alpha \zeta_\beta 
+ \hat\nabla_\beta \zeta_\alpha
\end{equation}
Notice that $\epsilon_{\alpha\beta}$ defines $\zeta_\alpha$ only up to a
transformation

\begin{equation}
\label {3.50}
\zeta_\alpha \rightarrow \zeta_\alpha + s_\alpha
\end{equation} 
where $s_\alpha$ is the generator of a symetrie of the Fermat quo-tensor,
i.e. a solution of the following equations:

\begin{equation}
\label{3.51}
\hat\nabla_\alpha \zeta_\beta + \hat\nabla_\beta \zeta_\alpha = 0
\end{equation}

We shall say that a deformation is static if

\begin{equation} \label {3.29} \delta u^\alpha = 0 \end{equation} Notice
that this does imply neither that the generator vector is zero nor that the
strain tensor is zero. Static deformations are the simplest and they are the
only ones that we shall consider in the following two sections.

An elastic test body is described by its energy-momentum tensor:

\begin{equation}
\label {3.14}
T^\alpha_\beta=\rho u^\alpha u_\beta -c^{-2}\pi^\alpha_\beta, \quad
\pi^\alpha_\beta u^\beta=0
\end{equation}
where $\rho$ is its density, $u^\alpha$ is its unit four-velocity field and
where $\pi_{\alpha\beta}$ is the tensor describing the stresses which are 
present as a result of the volume forces acting on the body and those acting
on its surface. The body is described also by its shape, and this supposes a
geometry of space to which to refer it. If no other forces than
gravitational ones act on the body then its energy-momentum tensor must be
conserved:

\begin{equation}
\label {3.15}
\nabla_\alpha T^\alpha_\beta=0,
\end{equation}  
otherwise the r-h-s would be equal to the volume applied forces. On the
surface of the body we must have

\begin{equation}
\label {3.16}
\pi_{\alpha\beta}n^\alpha = f_\beta
\end{equation}
where $n^\alpha$ is the normal to the surface and $f_\beta$ are the applied
forces on the surface. 

Projecting eqs. (\ref{3.15}) along $u^\alpha$ one gets

\begin{equation}
\label {3.17}
u^\alpha\partial_\alpha \rho+\frac12 \rho\Sigma =
{1 \over{2 c^2}}\pi^{\alpha\beta}\Sigma_{\alpha\beta}
\end{equation}
where $\Sigma_{\alpha\beta}$ and $\Sigma$ have been defined in (\ref{2.18}) 
and (\ref{2.22})

On the other hand projecting eqs. (\ref{3.15}) on the tangent hyperplane
orthogonal to $u^\alpha$ one gets

\begin{equation}
\label {3.20}
\hat D_\alpha \pi^\alpha_\beta = -\rho \Lambda_\beta
\end{equation}
where

\begin{equation}
\label {3.21}
\hat D_\alpha \equiv \hat\nabla_\alpha -c^{-2}\Lambda_\alpha
\end{equation}
with $\Lambda^\alpha$ and $\hat \nabla_\alpha$ already defined in 
(\ref{2.9}) and (\ref{2.23}).

Let us consider all space-time metrics for which systems of coordinates
exist such that the behavior of its components
when $c\rightarrow \infty$ is:

\begin{equation}
\label {3.35}
g_{00}=-1 + {2U(t,x^k)\over c^2}, \quad
g_{0i}={U_i(t,x^k)\over c^3}, \quad
g_{ij}=\delta_{ij}
\end{equation}
Since this behavior is preserved by the infinite-dimensional classic group
of rigid motions:

\begin{equation}
\label {3.36}
t^\prime=t, \quad x^{i^\prime} = R^{i^\prime}_j(t)(x^j+S^j(t))
\end{equation}
where the matrix $R$ is a rotation matrix, we shall call the behavior above
the classical limit, if it exists, of a space-time metric.
\,\footnote{The metric components
\ref{3.35} can be considered as defining the space-time metric of an exact
`extended Newtonian theory' invariant under the group of rigid motions
\ref{3.36}. This theory was presented in \cite{Bel0}}. At this limit
the Fermat tensor is, neglecting terms of order greater than $c^{-2}$,

\begin{equation}
\label {2.35}
\hat g_{ij} = \delta_{ij}
\end{equation} 
and at the same approximation we have

\begin{equation}
\label {2.36}
\Lambda_i=\partial_i U - \frac{1}{c^2}\partial_t U_i, \quad
\Sigma_{ij}=0
\end{equation}
Equations (\ref{3.17}) and (\ref{3.20}) become then

\begin{equation}
\label {3.38}
\partial_t \rho=0, \quad \partial_i \pi^i_j=-\rho\Lambda_i,
\end{equation}
the latter being the equilibrium equations of the body.

Let us assume that we have an elastic body which is isotropic and
homogeneous, then we shall say that this body has been statically deformed
with respect to some given frame of reference $\cal R$ if the 4-velocity of
each element of the body is tangent to $\cal R$ and if a static deformation
of the frame of reference can be found such that the stress tensor of the
body can be related to the strain tensor according to the relativistic
generalization of the Hooke's law

\begin{equation}
\label {3.30}
\epsilon_{\alpha\beta}={1 \over {3(3\lambda+2\mu)}}\pi\hat g_{\alpha\beta}+
{1\over{2\mu}}(\pi_{\alpha\beta}-\frac13 \hat g_{\alpha\beta}\pi), \quad 
\pi=\hat g^{\alpha\beta}\pi_{\alpha\beta}
\end{equation}
where $\lambda$ and $\mu$ are the Lam\'e's parameters of the body. Of course
not every congruence which is a frame of reference can be the motion of a
body which has been statically deformed. In the next section we shall
consider a case for which this is the case because the congruence is a
Killing congruence, and in section 5 we shall consider a case for which this
is also the case but only at a desired approximation.
 
Since the strain tensor $\epsilon_{\alpha\beta}$ has a well defined
structure one can derive from eqs. (\ref{3.26}) necessary integrability
conditions that this tensor, or the displacement vector $\zeta_\alpha$ will
have to satisfy. Because of Hooke's law the stress tensor will have to
satisfy also some supplementary equations besides eqs. (\ref{3.30}). We are
not going to detail here these supplementary equations but we shall remind
what they are at the classical limit because we shall refer to them in the
next section.

From the definition (\ref{3.26}) we have at this limit

\begin{equation}
\label {3.39}
\epsilon_{ij}=\partial_i \zeta_j + \partial_j \zeta_i
\end{equation}    
whose integrability conditions are

\begin{equation}
\label {3.40}
\partial_{ik} \epsilon_{jl} +\partial_{jl} \epsilon_{ik} 
-\partial_{il} \epsilon_{jk} -\partial_{jk} \epsilon_{il}=0
\end{equation}
But since the l-h-s of these equations has the geometrical structure of a
linearized Riemann tensor in three dimensions these equations are equivalent
to its contraction with $\delta^{jl}$, i.e.:

\begin{equation}
\label {3.41}
\triangle\epsilon_{ik}+\partial_{ik}\epsilon^s_s
-\partial_i(\partial_s\epsilon^s_k)-\partial_k(\partial_s\epsilon^s_i)=0
\end{equation}
and using Hooke's law and the equilibrium equations (\ref{3.38}) one obtains:

\begin{equation}
\label {3.42}
\triangle\pi_{ik}+ \frac{2(\mu-\lambda)}{2\mu+3\lambda}\partial_{ik}\pi=
\rho(\partial_i \Lambda_k + \partial_k \Lambda_i 
+ \frac{\lambda}{2\mu-\lambda}\partial_s\Lambda^s\delta_{ik})
\end{equation}
These equations are called the Beltrami-Michell equations (See for instance
\cite{Filonenko}). Their importance comes from the fact, known as the
fundamental theorem of the theory of elasticity theory in classical
mechanics, that joined to the equilibrium equations (\ref{3.38}) and the
conditions on the surface (\ref{3.16}) they determine a unique solution for
the stress tensor and, through Hooke's law, a unique solution for the strain
tensor. The displacement vector is then uniquely defined up to a rigid
motion. 

\section{Constant acceleration in Minkowski space-time}

Let us consider an homogeneous and isotropic test body. From the elasticity
point of view this body will be characterized by its density and its
Lam\'e's parameters. And from the geometrical point of view it will be
characterized by its shape, a concept that supposes a geometry of the space
on which it is embedded. This geometry, from our point of view, is that of
the quotient manifold ${\cal V}_3$ and the geometric ingredients defined in
(\ref{2.3})and (\ref{2.15}).

This section is meant to be a simple application of the relativistic theory
of static deformations. Therefore to keep things as simple as possible we
shall assume that the space-time is Minkowski's and that the body is a
cylindrical body at rest initially with respect to a galilean frame of
reference. We shall assume also that the basis of the cylinder is pushed
with a uniform force along a fixed direction.  We assume finally that after
some transient elastic oscillations, that we shall not consider here, the
motion of the body is the Killing congruence $u^\alpha$ obtained
constructing the irrotational Fermi congruence with origin any of the
world-lines of the basis on which acts the constant force.

Using a system of coordinates adapted to $u^\alpha$ such that

\begin{equation}
\label {4.4}
u^0 = (1+c^{-2}a_k x^k)^{-1}, \quad u^i = 0
\end{equation}
the line-element of Minkowski's space-time is (See for instance
\cite{Moller})

\begin{equation}
\label {4.1}
ds^2=-[1+c^{-2}a_i x^i]^2 c^2dt^2+\delta_{ij}dx^i dx^j
\end{equation}
where $a_i$ is the intrinsic acceleration of the world-line which has been
chosen as origin of the Fermi construction.  
The corresponding Fermat tensor is:

\begin{equation}
\label {4.12}
\hat g_{ij}=\delta_{ij}
\end{equation}
i.e., the space manifold ${\cal V}_3$ associated with the frame of reference
$\cal R$ is euclidean like in classical mechanics. The components of
$\Lambda_i$ and $\Sigma_{ij}$ are:

\begin{equation}
\label {4.5}
\Lambda_i = -a_i(1+c^{-2}a_k x^k)^{-1}, \quad \Sigma_{ij}=0 
\end{equation}
Note that each element of the
cylinder describes a world line with constant intrinsic acceleration but
this acceleration will depend on the distance to the basis of the cylinder.
Equations (\ref{3.17}) become in this example

\begin{equation}
\label {4.2}
\partial_t \rho = 0,
\end{equation}
and equations (\ref{3.20}) become

\begin{equation}
\label {4.6}
\partial_i \pi^i_j + c^{-2}(1+c^{-2}a_k x^k)^{-1}a_i\pi^i_j = 
(1+c^{-2}a_k x^k)^{-1}\rho a_j
\end{equation}
As we see, at the classical limit, these are the equilibrium equations

\begin{equation}
\label {4.15}
\partial_i \pi^i_j = \rho a_j
\end{equation}
of a body which moves with uniform constant acceleration. 

Let us assume now that the basis of the cylinder is parallel to the $x-y$
plane and it is being pushed in the positive $z$ direction.Taking into
account the symmetries of the problem and the fact that $a_1 =a_2 = 0$ it is
easy to find a particular solution to these equations satisfying the
boundary conditions (\ref{3.16}), namely:

\begin{equation}
\label {4.3}
\pi^3_3 = -(1+c^{-2}a z)^{-1}\rho a(l-z), \quad a=a_3
\end{equation}
$l$ being the height of the cylinder and the other components of the stress 
tensor being zero. At the classical limit we have:

\begin{equation}
\label {4.15b}
\pi^3_3 = -\rho a(l-z)
\end{equation}
which is the solution that leads to  the correct solution for the generator
vector $\zeta_i$. But in Special relativity (\ref{4.3}) it is not. In fact
from (\ref{3.26}) and (\ref{4.12}) it follows that the strain tensor is
again the same expression (\ref{3.39}) as in classical mechanics; therefore
the integrability conditions for these equations are the Beltrami-Michell
equations (\ref{3.42}) that we derived in the preceding section and a direct
substitution of expression (\ref{4.3}), the other components of the stress
tensor being zero, would prove that this tensor does not satisfy these eqs.
(\ref{3.42}). 

To solve the problem in this case we could look for a solution in terms of 
power series of $1/c^2$

\begin{equation}
\label {4.11}
\pi_{ij} = \pi^{(0)}_{ij} + 1/c^2 \pi^{(1)}_{ij}+ \cdots
\end{equation}
starting with the classical solution (\ref{4.15}). At each order of
approximation one has to solve then a classical problem of elasticity with
different volume forces.

\section{Weak plane gravitational waves}

Let us consider the line element of a weak plane gravitational wave that we
shall assume for simplicity to be rectilinearly polarized. To be more
precise, we consider the line element:

\begin{equation}
\label {5.1}
ds^2=-c^2dt^2 + (1+h)dx^2+(1-h)dy^2+dz^2, \quad h=h(u), \quad u=ct-z
\end{equation} 

This is actually the wave-zone expression of a particular approximate 
solution

\begin{equation}
\label {5.3}
g_{\alpha\beta}=\eta_{\alpha\beta}+h_{\alpha\beta}+\cdots
\end{equation}
of the linearized Einstein equations:

\begin{equation}
\label {5.2}
\Box \bar h_{\alpha\beta} = T_{\alpha\beta}, \quad 
\bar h_{\alpha\beta} = h_{\alpha\beta}-\frac12 h \eta_{\alpha\beta}, \quad
h=\eta^{\alpha\beta}h_{\alpha\beta}
\end{equation}
To obtain this solution it has been assumed that the source of this
gravitational field was confined in a bounded region of the space associated
with a galilean frame of reference of Minkowski space-time. Namely the
congruence $\cal R$ with parametric equations

\begin{equation}
\label {5.4}
x^i=const. 
\end{equation}
and unit tangent vector with components $u^0=1$, $u^i=0$. The concept of a
Galilean frame of reference is a well defined concept if the space-time is
Minkowski's but not for a space-time with a metric like (\ref{5.3}). What
one needs is a definition that generalizes the concept of a frame of
reference acceptable in any space-time. Our approach of section 2 to this
problem allows us to interpret the congruence defined by eqs. (\ref{5.4}) as
a frame of reference with respect to which the source of the field is at
rest, not because it is a galilean frame of reference of Minkowski
space-time but because it is, at the approximation being considered, a
quo-harmonic congruence which will reduce, in this particular case, to a
galilean frame of reference if $h$ in (\ref{5.1}) were zero. In fact the
geometrical objects associated with the congruence (\ref{5.4}) are the
following: the Fermat tensor is

\begin{equation}
\label {5.16}
\hat g_{ij}=g_{ij}=\delta_{ij}+\hat h_{ij}, \quad 
\hat h_{ij}=h(\delta_{1i}\delta_{1j}-\delta_{2i}\delta_{2j});
\end{equation}
the Newtonian field of forces is zero

\begin{equation}
\label {5.17}
\Lambda_i=0
\end{equation}
and the non zero components of the rate of 'deformation' field are

\begin{equation}
\label {5.18}
\Sigma_{11}= c\partial_0 h, \quad \Sigma_{22}=-c\partial_0 h
\end{equation}
From (\ref{5.16}) it follows that

\begin{equation}
\label {5.21}
\hat\triangle x^k = -\frac12\delta^{ij}\delta^{ks}
(\partial_i \hat h_{js} +\partial_j \hat h_{is}-\partial_s \hat h_{ij})
\end{equation}
which yields:

\begin{equation}
\label {5.22}
\hat\triangle x^1 = \partial_1 h=0, \quad
\hat\triangle x^2 = \partial_2 h=0, \quad
\hat\triangle x^3 = 0
\end{equation}
meaning that the congruence $\cal R$ is quo-harmonic.  At the desired
approximation we also have

\begin{equation}
\label {5.19}
\hat g^{11}=1-h, \quad \hat g^{22}=1+h, \quad \hat g^{33}=1
\end{equation}
the other components being zero. From these expressions we can see that the
relations (\ref{2.13}) are satisfied with

\begin{equation}
\label {5.20}
c_1=-\partial^2_0 h/\partial_0 h
\end{equation}
which means that the congruence has locally genericity 6.

Let us examine the problem of deciding how an elastic bar couples to a wave
like (\ref{5.1}). To do that we ask whether an elastic bar at rest or in
uniform motion with respect to the source of the wave can be statically
deformed with respect to the frame of reference $\cal R$, and if so under
which forces acting on its surface. To implement the idea that the bar is at
rest with respect to the source we have to accept that the motion of the bar
is itself described by the congruence $\cal R$. Taking into account the
smallness of the deviation with respect to the euclidean metric of the
Fermat tensor associated with $\cal R$, the smallness of the stress tensor,
and eqs. (\ref{5.17}) the equilibrium equations (\ref{3.20}) become here

\begin{equation}
\label {5.5}
\partial_i \pi^i_j=0 
\end{equation}
and the expression of the strain tensor	(\ref{3.26}) is the same as in the
classical limit

\begin{equation}
\label {5.46}
\epsilon_{ij}=\partial_i \zeta_j + \partial_j \zeta_i
\end{equation}
Since

\begin{equation}
\label {5.7}
\pi_{ij} =0
\end{equation}
is an obvious solution of the equilibrium equations (\ref{5.5}) which
satisfies the boundary conditions (\ref{3.16}) with zero applied forces and
since the integrability conditions of eqs. (\ref{5.46}) are also trivially
satisfied it follows from Hooke's law (\ref{3.30}) that $\epsilon_{ij}=0$
and that $\zeta_i$ is either zero or it defines an infinitesimal fixed rigid
motion. This result can be translated in words as follows: an elastic bar,
if left at rest in the wave zone, or in uniform motion with respect to a
source of gravitational radiation, will remain in the same state of motion
and no stresses will be induced at this approximation by the wave.

\section*{Acknowledgments}

It is a pleasure to acknowlege many stimulating and useful discussions about
the relativistic theory of elaticity with J. Mart\'{\i}n, J. Llosa and P.
Teyssandier.

\end{document}